\documentclass[prl, apl, aps4, twocolumn, superscriptaddress]{revtex4}
\usepackage{amsmath}
\usepackage{amsfonts}
\usepackage{amssymb}
\usepackage{graphics}
\usepackage{graphicx}
\usepackage{color}
\usepackage[applemac]{inputenc}
\usepackage[cyr]{aeguill}
\usepackage{psfrag}

\newcommand{\ii}{\text{i}}

\newcommand{\um}{\mu\text{m}}

\definecolor{green2}{rgb}{0,0.66,0} 

\begin{document}

\title{Manipulating bubbles with secondary Bjerknes forces}
\author{Maxime Lanoy}
\affiliation{Institut Langevin, ESPCI ParisTech, CNRS (UMR 7587), PSL Research University, Paris, France}
\affiliation{Laboratoire Mati\`ere et Syst\`emes Complexes, Universit\'e Paris-Diderot, CNRS (UMR 7057), Paris, France}

\author{Caroline Derec}
\affiliation{Laboratoire Mati\`ere et Syst\`emes Complexes, Universit\'e Paris-Diderot, CNRS (UMR 7057), Paris, France}

\author{Arnaud Tourin}
\affiliation{Institut Langevin, ESPCI ParisTech, CNRS (UMR 7587), PSL Research University, Paris, France}

\author{Valentin Leroy}
\affiliation{Laboratoire Mati\`ere et Syst\`emes Complexes, Universit\'e Paris-Diderot, CNRS (UMR 7057), Paris, France}


\date{\today}

\begin{abstract}
Gas bubbles in a sound field are submitted to a radiative force, known as the secondary Bjerknes force. We propose an original experimental setup that allows us to investigate in details this force between two bubbles, as a function of the sonication frequency, as well as the bubbles radii and distance. We report the observation of both attractive and, more interestingly, repulsive Bjerknes force, when the two bubbles are driven in antiphase. Our experiments show the importance of taking multiple scattering into account, which leads to a strong acoustic coupling of the bubbles when their radii are similar. Our setup demonstrates the accuracy of secondary Bjerknes forces for attracting or repealing a bubble, and could lead to new acoustic tools for non contact manipulation in microfluidic devices.
\end{abstract}

\maketitle

Microfluidics is a fast growing field that has motivated the development of non-contact methods for manipulating both fluids and particles within those fluids at the microscale. The manipulation is done through a force exerted directly on an object or by shaping the streamlines of the flow~\cite{fu1999microfabricated,franke2010surface,yasuda1996particle,bisceglia2013micro}. To this day the most popular tool for manipulating small objects is probably the optical tweezer~\cite{ashkin1970acceleration, ashkin1986observation}, which can be used to displace dielectric objects, or to pull them with a calibrated force and a nanometric resolution. An acoustic field can also be used to exert radiative forces on small objects and some applications to microfluidics have already been investigated ~\cite{baresch2014observation,tran2012fast,rabaud2011manipulation}. High frequency ultrasounds are usually required to focus acoustic energy at the desired scale.

In this Letter, we investigate the possibility of using ultrasound to manipulate sub-wavelength objects without limitation of resolution. At the large wavelength limit, a pressure wave exerts an instantaneous force $-V. \nabla P$ over an object with volume $V$. In general the pressure gradient takes alternatively positive and negative values and the net force thus averages to zero. The situation is however different when the volume of the object also changes with time, as is the case for a bubble in a stationary acoustic field: the resulting force, known as the primary Bjerknes force, drives the bubble towards the pressure nodes or antinodes~\cite{bjerknes1906fields,bjerknes1909kraftfelder,eller1968force}. When two, or more, bubbles are driven by an incident sound field, they scatter a secondary field which tends to cause another force: the secondary Bjerknes force.  

We report on an original setup where a fixed pulsating bubble is used as an actuator: by applying ultrasound, the trajectory of a bubble flowing in its vicinity is modified through the action of the secondary Bjerknes force. We carefully study the dependence of that force on the radii of the bubbles, their distance and the frequency of the exciting acoustic field. 

\begin{figure}[htb!]
    \centering
      \includegraphics[width=\linewidth]{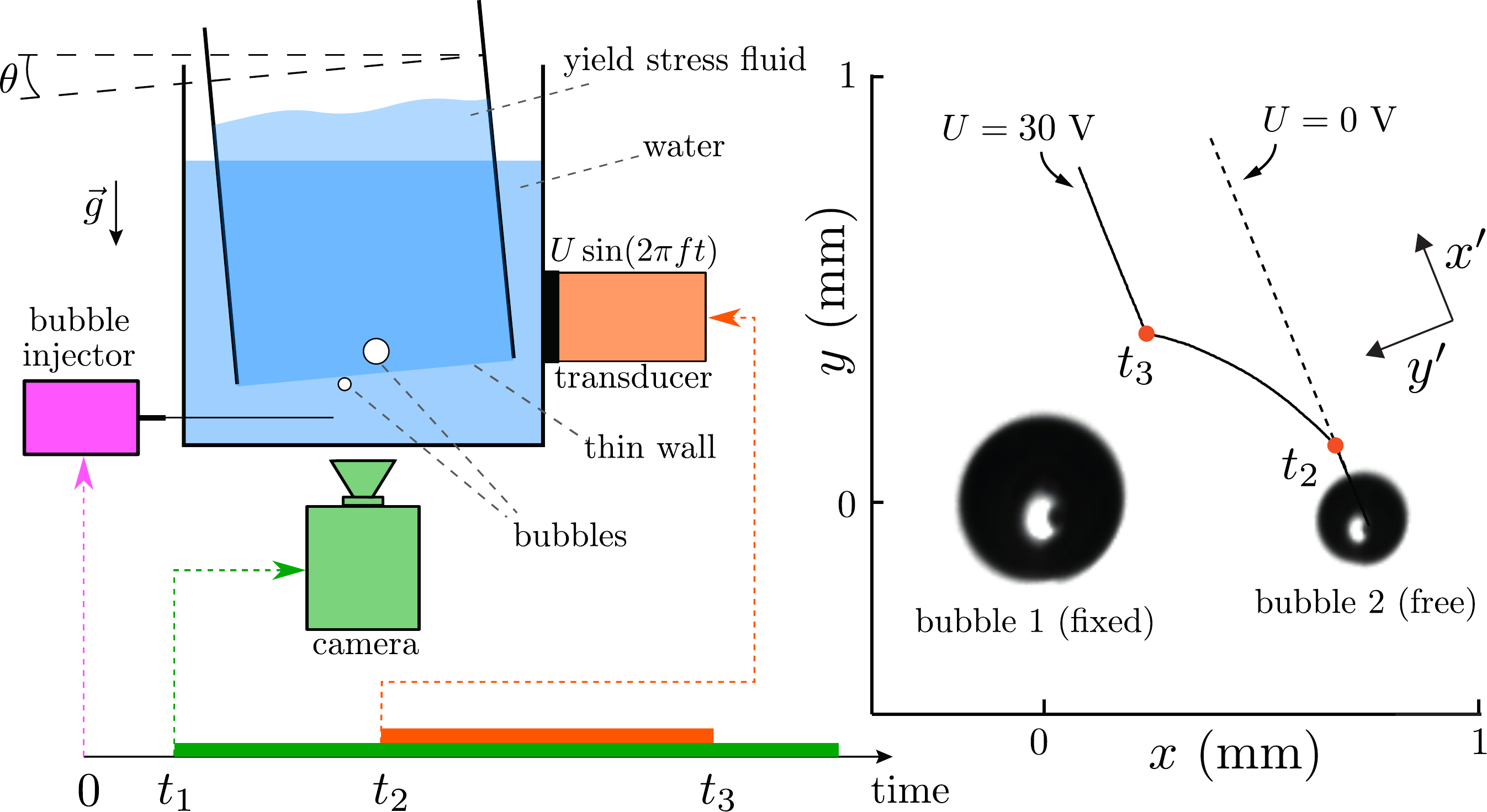}
    \caption{We investigate the secondary Bjerknes force at play between two bubbles by observing the trajectory of a free bubble (in water) under the influence of a fixed bubble (in a yield-stress fluid) when a pressure field is applied. The two bubbles are separated by a thin wall that is acoustically transparent. The pressure field is generated by a transducer driven by a voltage signal $U\sin(2\pi f t)$. Here we show two trajectories obtained with a $R_1=200\,\mu$m fixed bubble and a $R_2=115\,\mu$m free bubble: with no sonication ($U=0$), and with a $U=30\,$V signal at $f=14\,$kHz.}\label{setup}
\end{figure}

The setup is schematized in figure~\ref{setup}; it consists of two nested tanks, the largest of which is filled with water. The inner tank is filled with a yield-stress fluid and its bottom face is closed by a $3\,\um$-thick transparent wall (mylar), ensuring almost perfect acoustic impedance matching between both fluids in the range of frequencies we were investigating. It can be tilted by a small angle $\theta$ relatively to the water tank thanks to a microgoniometer. Prior to the experiment, a bubble (named bubble $1$) is injected and trapped at a fixed position inside the yield-stress fluid.
At time $0$, bubble $2$ is generated in water, and elevates under the action of buoyancy until it reaches the thin wall. Starting at time $t_1$, a high speed camera is used to observe the plane of the thin wall, with a depth of field large enough to image the two bubbles. The sizes of the bubbles can then be determined (with a $10\,\mu$m accuracy) and the trajectory of bubble $2$ followed. To investigate the effect of the acoustic field, a piezoelectric transducer (Panametrics X1021) is switched on between times $t_2$ and $t_3$, driven by a continuous sine wave with voltage $U$ and frequency $f$. As shown in Fig.~\ref{setup}, bubble $2$ has a straight trajectory,  due to the buoyancy force ($\theta \neq 0$), when no sound is applied ($U=0$). We note $x^\prime$ the direction followed by the bubble in this case and $y^\prime$ refers to the direct orthogonal direction within the wall plan.  A clear deviation of the trajectory toward bubble $1$ is observed when the bubbles are insonified ($U\neq 0$), suggesting that a secondary Bjerknes force is indeed at play. Note that, in order to observe secondary Bjerknes forces, one has to insure they dominate over primary forces. Two precautions were taken in that sense. First, the water tank was chosen sufficiently narrow ($1.5$ cm) to avoid cavity resonances. Second, bubble $1$ was injected sufficiently close to the thin wall to minimize the inter-bubble distance.

\begin{figure}[htb!]
    \centering
      \includegraphics[width=.9\linewidth]{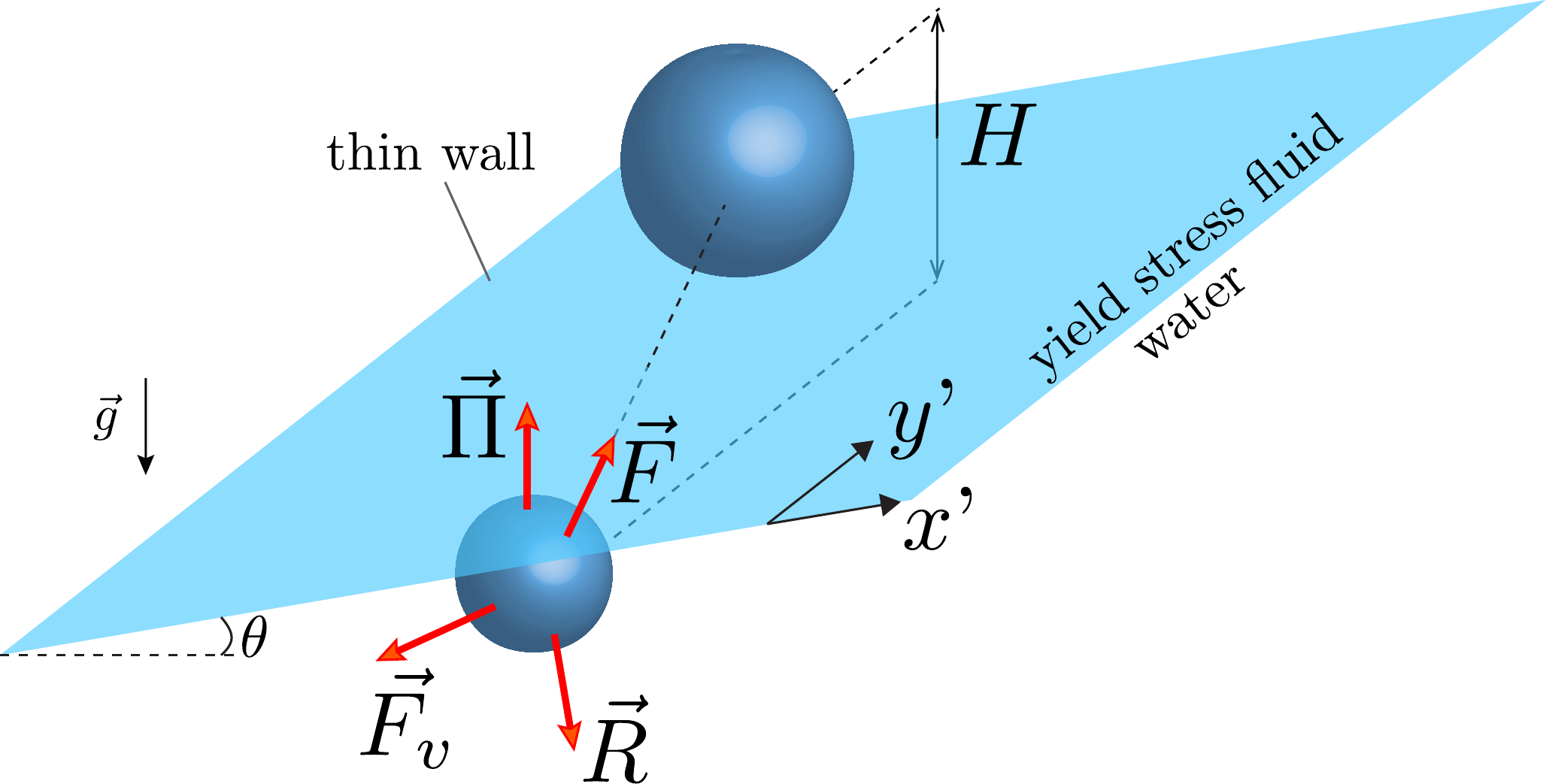}
    \caption{Bubble $2$, in water, is submitted to four forces: buoyancy force $\vec{\Pi}$, drag force $\vec{F}_v$, wall reaction $\vec{R}$ and Bjerknes force $\vec{F}$.}\label{fig3D}
\end{figure}

\begin{figure}[htb!]
    \centering
      \includegraphics[width=.70\linewidth]{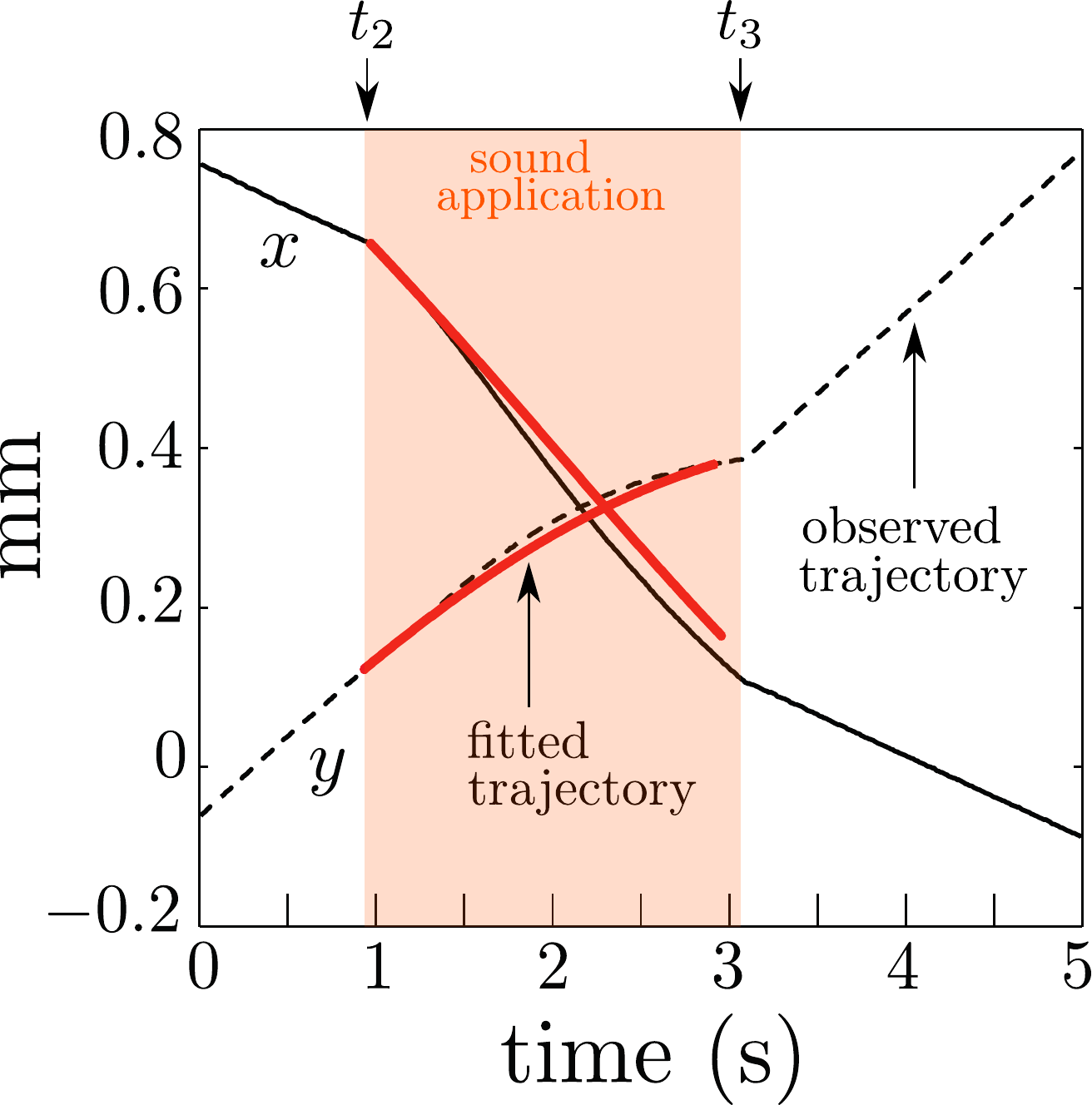}
    \caption{Position of bubble $2$ as a function of time, in the same condition as in Fig.~\ref{setup} ($U=30\,$V, $f=14\,$kHz). Using Eq.~(\ref{eqbilan2}) we can check that a central force inversely proportional to the squared distance is able to capture the observed trajectory, and determine the value of the prefactor $A$ of this force.}\label{fit}
\end{figure}

By analyzing trajectories such as the one depicted in Fig.~\ref{setup}, one can obtain quantitative information about the direction and intensity of the Bjerknes force at play, and compare it with the theoretical prediction. Figure~\ref{fig3D}  proposes a three dimensional view of the system, with an indication of the four forces that exert on bubble $2$.
The abrupt variations of the trajectory when the sonication starts and stops (see Fig.~\ref{setup}), suggest that inertia can be neglected. We assume (see eq.(~\ref{bjerknes})) that $F$ is a central force, inversely proportional to the square of the distance: $\vec{F}=-\alpha \vec{e}_r/r^2$ (where $\vec{e}_r$ is the unitary vector pointing from bubble $1$ to bubble $2$, and $r$ the distance between the bubbles), and that the drag force is of the form $\vec{F}_v=-\beta \vec{v}$ (where $\vec{v}$ is bubble $2$'s velocity). Projecting on the wall plane, we thus obtain the following dynamic equation: 
\begin{eqnarray}
-\alpha\frac{x'\vec{e_{x'}}+y'\vec{e_{y'}}}{(x'^2+y'^2+H^2)^{3/2}}+\Pi \sin\theta \vec{e_{x'}}-\beta\vec{v}=\vec{0}, \label{eqbilan}
\end{eqnarray}
where $H$ is the minimal distance between the bubbles ($H=R_1+R_2$ if bubble $1$ is touching the wall). Introducing bubble $2$'s velocity in the absence of sonication, when $\Pi \sin\theta = \beta v_0$, Eq.~(\ref{eqbilan}) can be rewritten
\begin{eqnarray}
\vec{v}(t)=v_0\vec{e}_{x'}-A\times \frac{x' \vec{e_{x'}}+y \vec{e_{y'}}}{(x'^2+y'^2+H^2)^{3/2}}, \label{eqbilan2}
\end{eqnarray}
where $A=\alpha/\beta$ is the only unknown parameter. By applying eq.~(\ref{eqbilan2}) on small successive time intervals, we can reconstruct the predicted trajectory. As shown in figure~\ref{fit}, it is possible to find a value of $A$ that gives a reasonable fitting of the observed trajectory.

\begin{figure}[htb!]
    \centering
   \includegraphics[width=.8\linewidth]{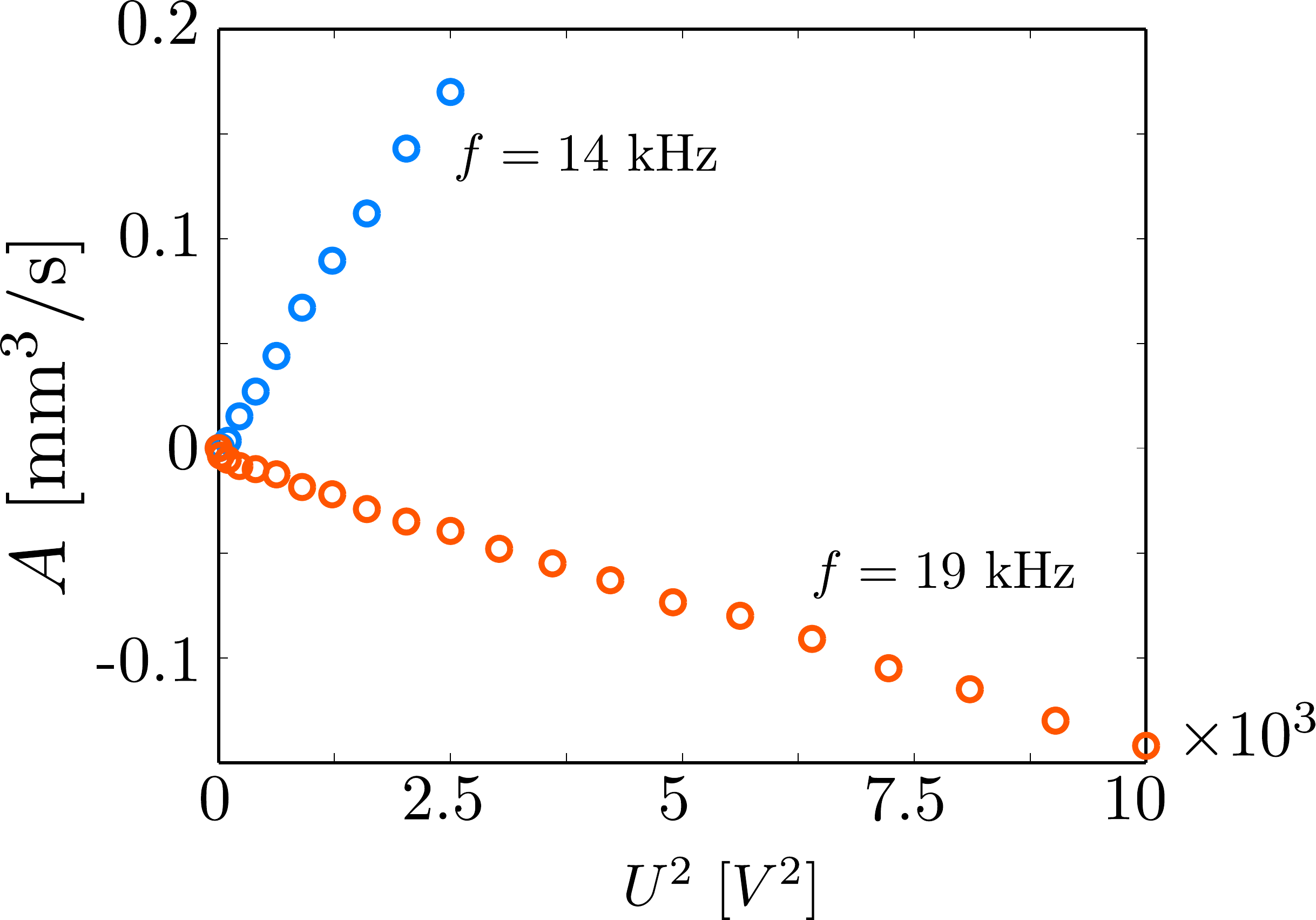}
    \caption{ Measured prefactor of the Bjerknes force as a function of the applied voltage squared, for two frequencies. Positive values of $A$ correspond to an attractive force, whereas negative ones indicate a repulsive force.}\label{couple1}
\end{figure}

For the couple of bubbles shown in Fig.~\ref{setup}, we acquired trajectories at different amplitudes of sonication, and for two frequencies. Figure~\ref{couple1} presents the values of $A$ we obtained with the fitting procedure. It appears that the force is attractive at $14\,$kHz and repulsive at $19\,$kHz, with an amplitude that scales as the square of the applied voltage in both cases. 

Althought theorical predictions~\cite{zabolotskaya1984interaction,doinikov1995mutual,ida2003alternative} have been available for some time, direct observations of a \emph{repulsive} secondary Bjerknes forces remain quite rare~\cite{yoshida2011experimental,barbat1999dynamics}, most of the experiments reporting only attractive forces~\cite{kazantsev1960motion,crum1975bjerknes,jiao2015experimental}. This arises from the difficulty of getting rid of the primary Bjerknes force in experiments.\\
When two bubbles are submitted to an oscillatory pressure field $p(t)=P \exp[-\ii\omega t]$, the force that bubble $1$ exerts on bubble $2$, as derived by Crum in $1975$~\cite{crum1975bjerknes}, is
\begin{equation}
\vec{F}_{1\to 2}=2\pi\rho (R_1 R_2 \omega)^2 \frac{\vec{r}_1-\vec{r}_2}{\lvert \vec{r}_1-\vec{r}_2 \rvert^3}\xi_1\xi_2\cos(\phi_1-\phi_2),
\label{bjerknes}
\end{equation}
where $\rho$ is the density of the liquid, $\vec{r}_n$ the position of bubble $n$, and where each bubble is oscillating with an instant radius $R_n+\xi_n \exp[-\ii(\omega t-\phi_n)]$. Eq.~(\ref{bjerknes}) predicts a force in $\xi^2$, which is consistent with the observed quadratic law as $\xi \propto P \propto U$. It also predicts a repulsive force when the bubbles oscillate in antiphase ($\phi_1-\phi_2=\pi$). In a naive picture of two uncoupled bubbles, bubbles oscillate as individual harmonic oscillators:
\begin{equation}
\xi e^{i\phi}=\frac{P}{\rho R_0} \times \frac{1}{\omega^2-\omega_M^2+\ii\omega^2\delta},
\label{xi}
\end{equation}
where $\omega_M$ is the Minnaert resonance angular frequency~\cite{minnaert}, and $\delta$ the damping rate. 
One thus expect the secondary Bjerknes force to be repulsive for frequencies between the two Minnaert frequencies of the bubbles, and attractive otherwise. For the two bubbles shown in Fig.~\ref{setup}, the Minnaert frequencies are $15\,$kHz and $25\,$kHz, which is consistent with the repulsion we observed at $19\,$kHz. 

\begin{figure}[htb!]
    \centering
      \includegraphics[width=.8\linewidth]{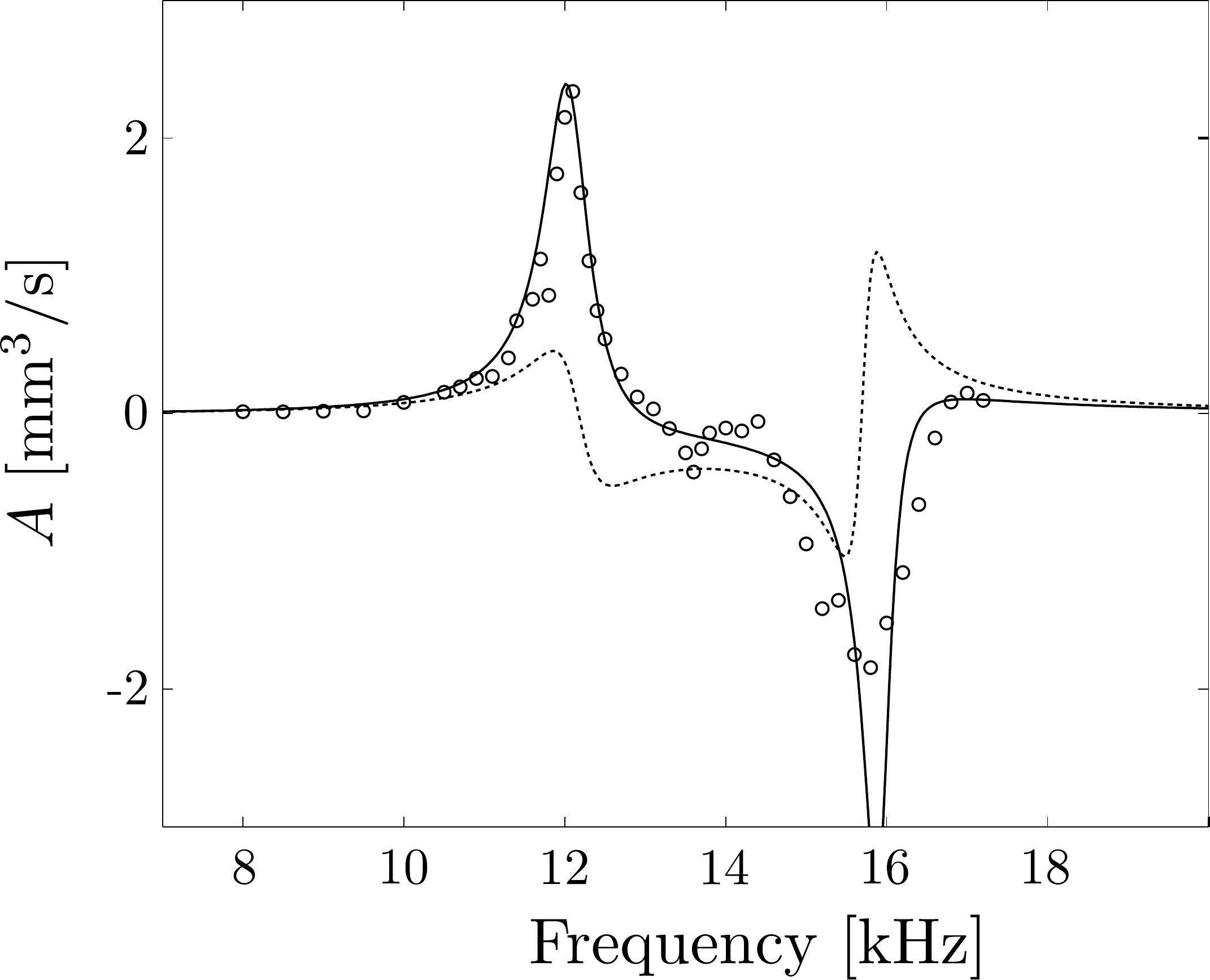}
    \caption{Symbols: measured prefactor $A$ as a function of frequency for $R_1=240\,\mu$m and $R_2=185\,\mu$m under an excitation of $U=60$ V. Lines: theoretical predictions without (dashed line) and with (solid line) coupling taken into account, assuming that the pressure generated by the transducer was constant with frequency.}\label{freq}
\end{figure}

Eq.~(\ref{xi}) is actually an approximation, valid only when the bubbles can be considered as uncoupled, \emph{i.e.} when the field scattered by bubble $1$ does not affect the response of bubble $2$. The effect of coupling  is stronger when the bubbles have similar radii, as can be evidenced by experiments with another couple of bubbles, with $R_{1}=240\,\mu$m and $R_{2}=185\,\mu$m ($R_1/R_2=1.30$, while it was $1.74$ in the previous case). We did a systematic observation of the trajectory of bubble $2$ as a function of the applied frequency. As shown in Fig.~\ref{freq}, $A$ presents a positive peak at the resonance frequency of the large bubble ($12\,$kHz), and a negative peak at the resonance frequency of the small bubble ($16\,$kHz). This is a quite different shape from what is predicted by the uncoupled model, in which the repulsion is expected in between the two resonances (see dashed curve in Fig.~\ref{freq}). A better model has to take the coupling into account, which consists in replacing, in (\ref{xi}), the applied pressure amplitude $P$ by the total pressure resulting from the multiple scattering from one bubble to each other. Practically, this only requires the inversion of a $2\times 2$ matrix, as detailed in previous works~\cite{leroy2005bubble,lanoy2015subwavelength}. When two bubbles are coupled they cannot be considered as two individual resonators. Instead, they form a two-level system with a symmetric mode, at a resonance close to the Minnaert frequency of the largest bubble, and an antisymmetric mode, at a resonance close to the Minnaert frequency of the smallest bubbles. Given the dependence of the Bjerknes force on the relative phase of the bubbles, it is natural to obtain an attractive force for the symmetric mode, and a repulsive force for the antisymmetric one. We show in Fig.~\ref{freq} that taking the coupling into account gives a correct prediction of the observed behavior. The experimental measurements of the bubble speed as a function of the tilt angle (not shown) lead to a value of $\beta=7\times10^{-6}$ kg.s$^{-1}$. At $f=12$kHz, we measure $A=2$mm$^3/$s and the corresponding magnitude for the secondary Bjerknes force is of $F_b=14$nN.

\begin{figure}[htb!]
    \centering
      \includegraphics[width=\linewidth]{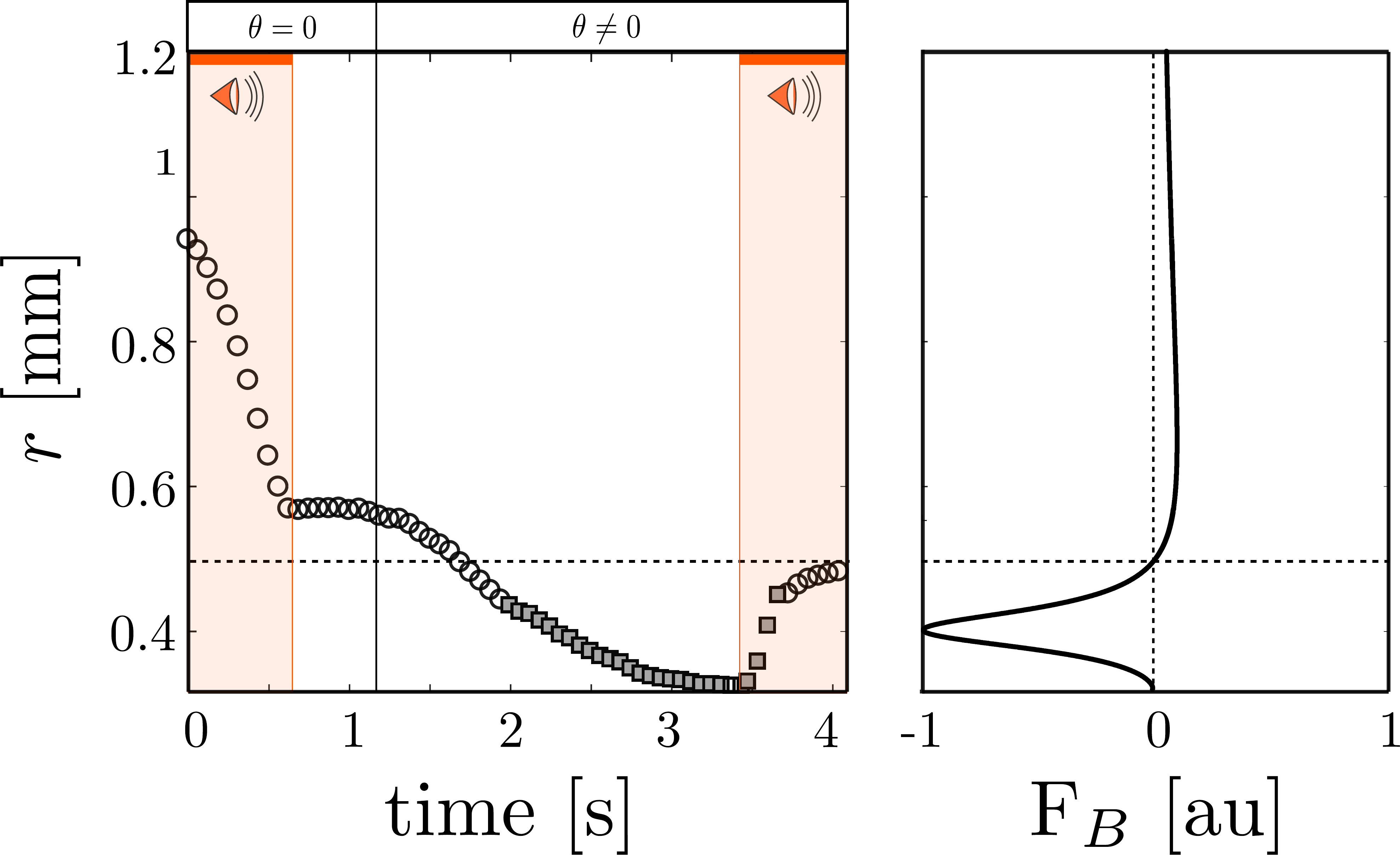}
    \caption{Left: experimental measurement of distance $r$ between bubbles as a function of time, for $R_1=151\,\mu$m and $R_2=127\,\mu$m (we suppose that $H=R_1+R_2$). Periods of tilting and sonication are highlighted on the graph.  Right: Analytical prediction of the secondary Bjerknes force given by eq.~\ref{bjerknes} with $\xi_1$ and $\xi_2$ calculated taking coupling into account. We adjust the value of radii $R_1$ and $R_2$ by setting the force sign inversion at the right interbubble distance and find $R_1=162\mu m$ and $R_2=132\mu m$ (consistent with our measurement uncertainty).}\label{sign}
\end{figure}

When coupling is strong, it can invalidate our approximation of a central force that depends only on the distance between the bubbles ($F=-\alpha /r^2$). Indeed, for strongly coupled bubbles, their oscillations are affected by their distance ($\xi_1$ and $\xi_2$ depends on $r$ in (\ref{bjerknes})). A striking illustration is given by an experiment in which bubbles are closer to each other, and more similar in size (two factors that enhance coupling). Let us consider two bubbles with $R_1=151\,\mu$m and $R_2=127\,\mu$m ($R_1/R_2=1.19$). Fig.~\ref{sign} shows how the distance between the bubbles changes in the following experiment. First, the angle of the tank is set to zero. When the sound field is applied, at a frequency of $25.5\,$kHz, bubble $2$ is attracted by bubble $1$. When the sound field is switched off, the bubble stops. The tank is then tilted to make bubble $2$ approach bubble $1$. The sound field is then switched on again, at the same frequency. This time, it results in a repulsion of the bubble. It clearly shows that the Bjerknes force can be attractive at long distance, and repulsive at short distance. This phenomenon is theoretically well documented by the works of Zabolotskaya~\cite{zabolotskaya1984interaction}, Doinikov~\cite{doinikov2001acoustic} and Ida~\cite{ida2003alternative} and has been observed under the form of trajectory oscillations by Barbat~\cite{barbat1999dynamics}. The right plot in Fig.~\ref{sign} shows that the theory indeed predicts an attractive force that becomes repulsive at short distance. This can be explained by a shift of the modes resonance frequencies. As the interbubble distance decreases, the frequency of the symmetric mode decreases, while the frequency of the antisymmetric mode increases. In our experiment, the antisymmetric peak is first at a frequency lower than $25.5\,$kHz, when bubbles are far appart; Bjerknes force is attractive. As the distance decreases, the frequency of the antisymmetric peak increases and eventually approaches $25.5\,$kHz, resulting in a repulsion when bubbles are close to each other.

Our original setup, with a fixed bubble in a yield-stress fluid and a free bubble in water, allowed us to study the secondary Bjerknes forces in detail. Repulsive Bjerknes forces, which have been seldom reported experimentally, were clearly observed and well predicted by the standard Crum formula, providing that the coupling was properly taken into account. Beyond the fundamental study, this device could be useful for microfluidic applications. Using several bubbles of different radii trapped in a yield-stress fluid (or in a soft solid), one could for example imagine guiding a free bubble toward different locations, depending on the frequency of the applied sound. Instead of being limited by the shape of the acoustic beam, the resolution would be given by the size of the bubbles, which act as local acoustic sources.
\\
\begin{acknowledgments}
We thank Cedric Poulain for his advice on bubble injection, and Mathieu Receveur for his decisive help on the design of the experimental setup. This work is supported by LABEX WIFI  (Laboratory of Excellence within the French Program ``Investments for the Future'') under references ANR-10-LABX-24 and ANR-10-IDEX-0001-02 PSL*. We thank Direction G\'en\'erale de l'Armement (DGA) for financial support to M.L. 
\end{acknowledgments}


\begin{thebibliography}{24}
\expandafter\ifx\csname natexlab\endcsname\relax\def\natexlab#1{#1}\fi
\expandafter\ifx\csname bibnamefont\endcsname\relax
  \def\bibnamefont#1{#1}\fi
\expandafter\ifx\csname bibfnamefont\endcsname\relax
  \def\bibfnamefont#1{#1}\fi
\expandafter\ifx\csname citenamefont\endcsname\relax
  \def\citenamefont#1{#1}\fi
\expandafter\ifx\csname url\endcsname\relax
  \def\url#1{\texttt{#1}}\fi
\expandafter\ifx\csname urlprefix\endcsname\relax\def\urlprefix{URL }\fi
\providecommand{\bibinfo}[2]{#2}
\providecommand{\eprint}[2][]{\url{#2}}

\bibitem[{\citenamefont{Fu et~al.}(1999)\citenamefont{Fu, Spence, Scherer,
  Arnold, and Quake}}]{fu1999microfabricated}
\bibinfo{author}{\bibfnamefont{A.~Y.} \bibnamefont{Fu}},
  \bibinfo{author}{\bibfnamefont{C.}~\bibnamefont{Spence}},
  \bibinfo{author}{\bibfnamefont{A.}~\bibnamefont{Scherer}},
  \bibinfo{author}{\bibfnamefont{F.~H.} \bibnamefont{Arnold}},
  \bibnamefont{and} \bibinfo{author}{\bibfnamefont{S.~R.} \bibnamefont{Quake}},
  \bibinfo{journal}{Nature Biotechnology} \textbf{\bibinfo{volume}{17}},
  \bibinfo{pages}{1109} (\bibinfo{year}{1999}).

\bibitem[{\citenamefont{Franke et~al.}(2010)\citenamefont{Franke,
  Braunm{\"u}ller, Schmid, Wixforth, and Weitz}}]{franke2010surface}
\bibinfo{author}{\bibfnamefont{T.}~\bibnamefont{Franke}},
  \bibinfo{author}{\bibfnamefont{S.}~\bibnamefont{Braunm{\"u}ller}},
  \bibinfo{author}{\bibfnamefont{L.}~\bibnamefont{Schmid}},
  \bibinfo{author}{\bibfnamefont{A.}~\bibnamefont{Wixforth}}, \bibnamefont{and}
  \bibinfo{author}{\bibfnamefont{D.}~\bibnamefont{Weitz}},
  \bibinfo{journal}{Lab on a Chip} \textbf{\bibinfo{volume}{10}},
  \bibinfo{pages}{789} (\bibinfo{year}{2010}).

\bibitem[{\citenamefont{Yasuda et~al.}(1996)\citenamefont{Yasuda, Umemura, and
  Takeda}}]{yasuda1996particle}
\bibinfo{author}{\bibfnamefont{K.}~\bibnamefont{Yasuda}},
  \bibinfo{author}{\bibfnamefont{S.-i.} \bibnamefont{Umemura}},
  \bibnamefont{and} \bibinfo{author}{\bibfnamefont{K.}~\bibnamefont{Takeda}},
  \bibinfo{journal}{The Journal of the Acoustical Society of America}
  \textbf{\bibinfo{volume}{99}}, \bibinfo{pages}{1965} (\bibinfo{year}{1996}).

\bibitem[{\citenamefont{Bisceglia et~al.}(2013)\citenamefont{Bisceglia,
  Cubizolles, Mallard, Vinet, Francais, and Le~Pioufle}}]{bisceglia2013micro}
\bibinfo{author}{\bibfnamefont{E.}~\bibnamefont{Bisceglia}},
  \bibinfo{author}{\bibfnamefont{M.}~\bibnamefont{Cubizolles}},
  \bibinfo{author}{\bibfnamefont{F.}~\bibnamefont{Mallard}},
  \bibinfo{author}{\bibfnamefont{F.}~\bibnamefont{Vinet}},
  \bibinfo{author}{\bibfnamefont{O.}~\bibnamefont{Francais}}, \bibnamefont{and}
  \bibinfo{author}{\bibfnamefont{B.}~\bibnamefont{Le~Pioufle}},
  \bibinfo{journal}{Lab on a Chip} \textbf{\bibinfo{volume}{13}},
  \bibinfo{pages}{901} (\bibinfo{year}{2013}).

\bibitem[{\citenamefont{Ashkin}(1970)}]{ashkin1970acceleration}
\bibinfo{author}{\bibfnamefont{A.}~\bibnamefont{Ashkin}},
  \bibinfo{journal}{Physical Review Letters} \textbf{\bibinfo{volume}{24}},
  \bibinfo{pages}{156} (\bibinfo{year}{1970}).

\bibitem[{\citenamefont{Ashkin et~al.}(1986)\citenamefont{Ashkin, Dziedzic,
  Bjorkholm, and Chu}}]{ashkin1986observation}
\bibinfo{author}{\bibfnamefont{A.}~\bibnamefont{Ashkin}},
  \bibinfo{author}{\bibfnamefont{J.}~\bibnamefont{Dziedzic}},
  \bibinfo{author}{\bibfnamefont{J.}~\bibnamefont{Bjorkholm}},
  \bibnamefont{and} \bibinfo{author}{\bibfnamefont{S.}~\bibnamefont{Chu}},
  \bibinfo{journal}{Optics Letters} \textbf{\bibinfo{volume}{11}},
  \bibinfo{pages}{288} (\bibinfo{year}{1986}).

\bibitem[{\citenamefont{Baresch et~al.}(2014)\citenamefont{Baresch, Thomas, and
  Marchiano}}]{baresch2014observation}
\bibinfo{author}{\bibfnamefont{D.}~\bibnamefont{Baresch}},
  \bibinfo{author}{\bibfnamefont{J.-L.} \bibnamefont{Thomas}},
  \bibnamefont{and}
  \bibinfo{author}{\bibfnamefont{R.}~\bibnamefont{Marchiano}},
  \bibinfo{journal}{arXiv preprint arXiv:1411.1912}  (\bibinfo{year}{2014}).

\bibitem[{\citenamefont{Tran et~al.}(2012)\citenamefont{Tran, Marmottant, and
  Thibault}}]{tran2012fast}
\bibinfo{author}{\bibfnamefont{S.}~\bibnamefont{Tran}},
  \bibinfo{author}{\bibfnamefont{P.}~\bibnamefont{Marmottant}},
  \bibnamefont{and} \bibinfo{author}{\bibfnamefont{P.}~\bibnamefont{Thibault}},
  \bibinfo{journal}{Applied Physics Letters} \textbf{\bibinfo{volume}{101}},
  \bibinfo{pages}{114103} (\bibinfo{year}{2012}).

\bibitem[{\citenamefont{Rabaud et~al.}(2011)\citenamefont{Rabaud, Thibault,
  Raven, Hugon, Lacot, and Marmottant}}]{rabaud2011manipulation}
\bibinfo{author}{\bibfnamefont{D.}~\bibnamefont{Rabaud}},
  \bibinfo{author}{\bibfnamefont{P.}~\bibnamefont{Thibault}},
  \bibinfo{author}{\bibfnamefont{J.-P.} \bibnamefont{Raven}},
  \bibinfo{author}{\bibfnamefont{O.}~\bibnamefont{Hugon}},
  \bibinfo{author}{\bibfnamefont{E.}~\bibnamefont{Lacot}}, \bibnamefont{and}
  \bibinfo{author}{\bibfnamefont{P.}~\bibnamefont{Marmottant}},
  \bibinfo{journal}{Physics of Fluids}
  \textbf{\bibinfo{volume}{23}}, \bibinfo{pages}{042003}
  (\bibinfo{year}{2011}).

\bibitem[{\citenamefont{Bjerknes}(1906)}]{bjerknes1906fields}
\bibinfo{author}{\bibfnamefont{V.~F.~K.} \bibnamefont{Bjerknes}}
  (\bibinfo{year}{1906}).

\bibitem[{\citenamefont{Bjerknes}(1909)}]{bjerknes1909kraftfelder}
\bibinfo{author}{\bibfnamefont{V.}~\bibnamefont{Bjerknes}},
  \emph{\bibinfo{title}{Die Kraftfelder}}, \bibinfo{number}{28}
  (\bibinfo{publisher}{F. Vieweg}, \bibinfo{year}{1909}).

\bibitem[{\citenamefont{Eller}(1968)}]{eller1968force}
\bibinfo{author}{\bibfnamefont{A.}~\bibnamefont{Eller}}, \bibinfo{journal}{The
  Journal of the Acoustical Society of America} \textbf{\bibinfo{volume}{43}},
  \bibinfo{pages}{170} (\bibinfo{year}{1968}).

\bibitem[{\citenamefont{Zabolotskaya}(1984)}]{zabolotskaya1984interaction}
\bibinfo{author}{\bibfnamefont{E.}~\bibnamefont{Zabolotskaya}},
  \bibinfo{journal}{Soviet Physics Acoustics}
  \textbf{\bibinfo{volume}{30}}, \bibinfo{pages}{365} (\bibinfo{year}{1984}).

\bibitem[{\citenamefont{Doinikov and Zavtrak}(1995)}]{doinikov1995mutual}
\bibinfo{author}{\bibfnamefont{A.}~\bibnamefont{Doinikov}} \bibnamefont{and}
  \bibinfo{author}{\bibfnamefont{S.}~\bibnamefont{Zavtrak}},
  \bibinfo{journal}{Physics of Fluids}
  \textbf{\bibinfo{volume}{7}}, \bibinfo{pages}{1923} (\bibinfo{year}{1995}).

\bibitem[{\citenamefont{Ida}(2003)}]{ida2003alternative}
\bibinfo{author}{\bibfnamefont{M.}~\bibnamefont{Ida}},
  \bibinfo{journal}{Physical Review E} \textbf{\bibinfo{volume}{67}},
  \bibinfo{pages}{056617} (\bibinfo{year}{2003}).

\bibitem[{\citenamefont{Yoshida et~al.}(2011)\citenamefont{Yoshida, Fujikawa,
  and Watanabe}}]{yoshida2011experimental}
\bibinfo{author}{\bibfnamefont{K.}~\bibnamefont{Yoshida}},
  \bibinfo{author}{\bibfnamefont{T.}~\bibnamefont{Fujikawa}}, \bibnamefont{and}
  \bibinfo{author}{\bibfnamefont{Y.}~\bibnamefont{Watanabe}},
  \bibinfo{journal}{The Journal of the Acoustical Society of America}
  \textbf{\bibinfo{volume}{130}}, \bibinfo{pages}{135} (\bibinfo{year}{2011}).

\bibitem[{\citenamefont{Barbat et~al.}(1999)\citenamefont{Barbat, Ashgriz, and
  Liu}}]{barbat1999dynamics}
\bibinfo{author}{\bibfnamefont{T.}~\bibnamefont{Barbat}},
  \bibinfo{author}{\bibfnamefont{N.}~\bibnamefont{Ashgriz}}, \bibnamefont{and}
  \bibinfo{author}{\bibfnamefont{C.-S.} \bibnamefont{Liu}},
  \bibinfo{journal}{Journal of Fluid Mechanics} \textbf{\bibinfo{volume}{389}},
  \bibinfo{pages}{137} (\bibinfo{year}{1999}).

\bibitem[{\citenamefont{Kazantsev}(1960)}]{kazantsev1960motion}
\bibinfo{author}{\bibfnamefont{V.}~\bibnamefont{Kazantsev}}, in
  \emph{\bibinfo{booktitle}{Soviet Physics Doklady}} (\bibinfo{year}{1960}),
  vol.~\bibinfo{volume}{4}, p. \bibinfo{pages}{1250}.

\bibitem[{\citenamefont{Crum}(1975)}]{crum1975bjerknes}
\bibinfo{author}{\bibfnamefont{L.~A.} \bibnamefont{Crum}},
  \bibinfo{journal}{The Journal of the Acoustical Society of America}
  \textbf{\bibinfo{volume}{57}}, \bibinfo{pages}{1363} (\bibinfo{year}{1975}).

\bibitem[{\citenamefont{Jiao et~al.}(2015)\citenamefont{Jiao, He, Kentish,
  Ashokkumar, Manasseh, and Lee}}]{jiao2015experimental}
\bibinfo{author}{\bibfnamefont{J.}~\bibnamefont{Jiao}},
  \bibinfo{author}{\bibfnamefont{Y.}~\bibnamefont{He}},
  \bibinfo{author}{\bibfnamefont{S.~E.} \bibnamefont{Kentish}},
  \bibinfo{author}{\bibfnamefont{M.}~\bibnamefont{Ashokkumar}},
  \bibinfo{author}{\bibfnamefont{R.}~\bibnamefont{Manasseh}}, \bibnamefont{and}
  \bibinfo{author}{\bibfnamefont{J.}~\bibnamefont{Lee}},
  \bibinfo{journal}{Ultrasonics} \textbf{\bibinfo{volume}{58}},
  \bibinfo{pages}{35} (\bibinfo{year}{2015}).

\bibitem[{\citenamefont{Minnaert}(1933)}]{minnaert}
\bibinfo{author}{\bibfnamefont{M.}~\bibnamefont{Minnaert}},
  \bibinfo{journal}{The London, Edinburgh, and Dublin Philosophical Magazine
  and Journal of Science} \textbf{\bibinfo{volume}{16}}, \bibinfo{pages}{235}
  (\bibinfo{year}{1933}).

\bibitem[{\citenamefont{Leroy et~al.}(2005)\citenamefont{Leroy, Devaud,
  Hocquet, and Bacri}}]{leroy2005bubble}
\bibinfo{author}{\bibfnamefont{V.}~\bibnamefont{Leroy}},
  \bibinfo{author}{\bibfnamefont{M.}~\bibnamefont{Devaud}},
  \bibinfo{author}{\bibfnamefont{T.}~\bibnamefont{Hocquet}}, \bibnamefont{and}
  \bibinfo{author}{\bibfnamefont{J.-C.} \bibnamefont{Bacri}},
  \bibinfo{journal}{The European Physical Journal E}
  \textbf{\bibinfo{volume}{17}}, \bibinfo{pages}{189} (\bibinfo{year}{2005}).

\bibitem[{\citenamefont{Lanoy et~al.}(2015)\citenamefont{Lanoy, Pierrat,
  Lemoult, Fink, Leroy, and Tourin}}]{lanoy2015subwavelength}
\bibinfo{author}{\bibfnamefont{M.}~\bibnamefont{Lanoy}},
  \bibinfo{author}{\bibfnamefont{R.}~\bibnamefont{Pierrat}},
  \bibinfo{author}{\bibfnamefont{F.}~\bibnamefont{Lemoult}},
  \bibinfo{author}{\bibfnamefont{M.}~\bibnamefont{Fink}},
  \bibinfo{author}{\bibfnamefont{V.}~\bibnamefont{Leroy}}, \bibnamefont{and}
  \bibinfo{author}{\bibfnamefont{A.}~\bibnamefont{Tourin}},
  \bibinfo{journal}{Phys. Rev. B} \textbf{\bibinfo{volume}{91}},
  \bibinfo{pages}{224202} (\bibinfo{year}{2015}),

\bibitem[{\citenamefont{Doinikov}(2001)}]{doinikov2001acoustic}
\bibinfo{author}{\bibfnamefont{A.~A.} \bibnamefont{Doinikov}},
  \bibinfo{journal}{Journal of Fluid Mechanics} \textbf{\bibinfo{volume}{444}},
  \bibinfo{pages}{1} (\bibinfo{year}{2001}).

\end{thebibliography}


\end{document}